\documentclass[11pt]{amsart}
\usepackage{geometry}                % See geometry.pdf to learn the layout options. There are lots.
\usepackage{graphicx}
\usepackage{mathrsfs}
\usepackage{amsmath,amsfonts,amssymb,amsthm}
\usepackage{epstopdf,natbib}
\usepackage[colorlinks,citecolor=blue,urlcolor=blue]{hyperref}
\usepackage{tikz}
\usetikzlibrary{positioning}
\newtheorem{theorem}{Theorem}[section]

\newtheorem{proposition}[theorem]{Proposition}

\theoremstyle{definition}

\theoremstyle{remark}

\newcommand{\QED}{\ifhmode\unskip\nobreak\fi\quad {\rm Q.E.D.}} % QED

\newcommand{\indep}{{\;\bot\!\!\!\!\!\!\bot\;}}

\title[Dependence of Bayesian Model Selection on
Irrelevant Alternatives]{The Dependence of Routine Bayesian Model Selection Methods on
Irrelevant Alternatives}
\author{Piotr Zwiernik}
\address{Department of Mathematics and Computer Science\\
TU Eindhoven\\PO Box 513 \\
5600 MB, Eindhoven\\ 
The Netherlands  }
\curraddr{}
\email{piotr.zwiernik@gmail.com}
\thanks{The first author was supported by Jan Draisma's Vidi grant from the Netherlands Organisation for ScientiÞc Research (NWO)}
\author{Jim Q. Smith}
\address{Department of Statistics\\
University of Warwick\\
Coventry\\
CV4 7AL\\
UK}
\curraddr{}
\email{j.q.smith@warwick.ac.uk}
\thanks{}
\begin{document}
\maketitle
\begin{abstract}
Bayesian methods - either based on Bayes Factors or BIC - are now widely
used for model selection. One property that might reasonably be demanded of
any model selection method is that if a model ${M}_{1}$ is preferred to a
model ${M}_{0}$, when these two models are expressed as members of one model
class $\mathbb{M}$, this preference is preserved when they are embedded
in a different class $\mathbb{M}'$. However, we illustrate in this
paper that with the usual implementation of these common Bayesian procedures this property does
not hold true even approximately. We therefore contend that to use these
methods it is first necessary for there to exist a "natural" embedding
class. We argue that in any context like the one illustrated in  our running example of Bayesian
model selection of binary phylogenetic trees there is no such embedding.
\end{abstract}

\section{Introduction}

Bayesian method such as Bayes Factor (BF) (for example \cite{denison2002bayesian}),
 or ones based on the Bayesian Information Criterion (BIC) \cite%
{schwarz1978edm} are now widespread in statistical analysis. In this paper we show that
a disadvantage of these approaches, in the way these are routinely used, is that they can give rise to an awkward
inferential ambiguity, which we later argue for some problems can never be resolved.  

Selecting a model can be seen as a special case of providing a preference
order over a number of options, which we assume to be finite. A long time ago
\cite{nash1950} and \cite{thomas84} argue that a particular property - restated below in terms that apply for model selection  - is essential for preference orderings used to identify an optimal choice. This states that if $M_{1}$ is preferred to $M_{0}$ (written $%
M_{1}\succ M_{0}$) in the model class $\mathbb{M}$ then $M_{1}\succ M_{0}$ also in $\mathbb{M%
}':=\mathbb{M }\cup\mathbb{M}^{+}$ whenever $M_{1}\succ
M^{+}$ for all $M^{+}\in \mathbb{M}^{+}$. So in particular if $M^{*}$ is a best model in $\mathbb{M}$ and we extend our selection to contain other models all of which are worse than $M^{*}$, then $M^{*}$ remains a best model in the larger set of models. This natural property is called \textit{%
independence of irrelevant alternatives} (IIA).

If IIA does not hold, then, whether or not we include poor fitting models in the selection set will influence what model we label as ``best''. Why should the
choice of this model be
influenced by inclusion or exclusion of other candidate models later
discovered to be poor representatives of the underlying data generating
process? Surely \textit{any} routine method of Bayesian Model selection: for example such as those reviewed by \cite{bernardosmith94} and \cite{keyetal1999} - even ones assuming that all models were wrong - should be expected not to violate IIA. 

Happily most applications of a
naive Bayes Factor (BF) model selection satisfy the property of IIA in
the following sense. Let $\mathbf{X}$ be the random vector of
observations over which models are selected, taking values $\mathbf{x}$
in the sample space $\mathcal{X}\subseteq \mathbb{R}^{m}$.  For simplicity
assume that all predictive densities associated with these models are
bounded and strictly positive over their shared support. Let $M_{0}$ be a
reference model with predictive density $p_{0}(\mathbf{x})$. Let $M_{%
{\phi}}$ be another candidate model with predictive density $p_{%
{\phi}}(\mathbf{x})$ on $\mathbf{x}\in \mathcal{X}$ and
a prior odds relative to $M_{0}$ of $\lambda _{{\phi}}$, where $%
\left( p_{{\phi}}(\mathbf{x}),\lambda _{{\phi}%
}\right) $ are functions of ${\phi}$ alone and in particular not
the predictive density and prior model probability $\left( p_{\mathbf{%
\phi }'}(\mathbf{x}),\lambda _{{\phi}'}\right) $ for any other candidate model $M_{{\phi}'}.$ 
Then setting $\lambda _{0}=1$, we note immediately that $M_{\mathbf{\phi 
}_{2}}\succeq M_{\phi_{1}}$ if and only if for the value of $%
\mathbf{x}$ we observe 
\begin{equation*}
\log p_{{\phi}_{2}}(\mathbf{x})+\log\lambda_{\phi_{2}}\geq \log p_{{\phi}_{1}}(\mathbf{x})+\log\lambda_{
{\phi}_{1}}.
\end{equation*}
Clearly this preference is therefore unaffected by the values of $\left( p_{{\phi}'}(\mathbf{x}),\lambda _{{\phi}'}\right)$ for $M_{{\phi}'}$ which may or may
not be contained in the selection class. 

Our problems begin when the class of models $\mathbb{M}$ over which
selection takes place is extremely large. Then the necessary task of
carefully and individually choosing separate prior distributions over the
hyperparameters of each candidate model is clearly infeasible. We are therefore forced
to reference our choice of prior density over the parameters of each
candidate so that in some sense these are consistent with each other. In this situation $(p_{\phi}(\mathbf{x}),\lambda_{\phi})$ is highly related with $(p_{\phi'}(\mathbf{x}),\lambda_{\phi'})$ for other candidate models and the BF may break the IIA property. 

For example this is exactly what happens when selecting over the class of Bayesian networks (BNs). In this case \cite{heckerman1995lbn}
introduced an additional demand that the prior densities  satisfied \textit{parameter modularity}. Here when parts of the structure of two of
these multivariate models coincide, then the priors over this
shared structure in these two models are assumed to be the same. This assumption, and others like it, not only makes the assignment of priors across a large model class feasible but also makes it possible to use greedy search algorithms to efficiently search the space for the best candidate model.  

However, these methods come at a price. Because prior densities are chosen to match those given within the model class, the choice of class itself can affect the inference and in particular IIA can be violated. Of course occasionally, in simple applications, there are compelling
reasons why a particular model class should be used. Then the violation of
IIA is not a problem. However, the choice of embedding class is often chosen
for convenience or convention rather than for some phenomenological reason
associated with the modelled process. It is in these circumstances that the
violation of IIA gives rise to poor inference. 

One such setting occurs when the modeler must decide whether or not to
include variables which usefully explain the process but cannot be
observed. Within the Bayesian paradigm the fact that these variables are not
observed does not matter in any formal sense: the score will simply be the
log-marginal density of the observed variables where we marginalize over the
missing ones. Thus this missingness causes no methodological problems. Indeed, if these integrations cannot be executed in closed form then their score can
be calculated in a straightforward manner using now standard numerical
techniques. The problem is rather that two statistically equivalent models
will be treated differently depending on whether or not the variables
representing these underlying causes are included. 

To be more concrete we consider the class of \textit{phylogenetic tree models}. The evolution of a collection of different extant species is typically represented by a Bayesian network on a directed tree, possibly with some additional constraints on the parameter vector (see for example \cite{semple2003pol}, \cite{yang2006cme}). The extant species are represented by the
leaves of a tree, whose interior vertices label extinct ancestors.  The model is usually depicted as a tree with edge lengths like for example in Figure \ref{fig:tree}. The topology of the tree represents the underlying graphical model. The lengths are functions of the conditional probabilities parametrising the model and in the phylogenetic context they give a measure of the phylogenetic distance between two species. 

\tikzstyle{leaf}=[circle,fill=black,minimum size=5pt,inner sep=0pt] %
\tikzstyle{vertex}=[circle,fill=yellow,draw,minimum size=5pt,inner sep=0pt] 
\begin{figure}[h]
\centering
\begin{tikzpicture}
    \node[vertex] (0) at (0,4)   {};
    \node[vertex] (1a) at (-1,3)   {};
    \node[vertex] (2a) at (1.5,2.3)   {};
    \node[vertex] (1b) at (-.7,0.7)   {};
    
    \node[leaf] (1) at (-2,0)  [label=below:$1$] {};
    \node[leaf] (2) at (-1,0)  [label=below:$2$] {};
    \node[leaf] (3) at (-0.3,0)  [label=below:$3$] {};
    \node[leaf] (4) at (1,0)  [label=below:$4$] {};
    \node[leaf] (5) at (2,0)  [label=below:$5$] {};
    \draw[<-,-latex,line width=.2mm] (0) to (1a);
    \draw[<-,-latex,line width=.2mm] (0) to (2a);
    \draw[<-,-latex,line width=.2mm] (1a) to (1);
    \draw[<-,-latex,line width=.2mm] (1a) to (1b);
    \draw[<-,-latex,line width=.2mm] (1b) to (2);
    \draw[<-,-latex,line width=.2mm] (1b) to (3);
    \draw[<-,-latex,line width=.2mm] (2a) to (4);
    \draw[<-,-latex,line width=.2mm] (2a) to (5);
  \end{tikzpicture}\qquad\qquad
\caption{A simple directed phylogenetic tree.}
\label{fig:tree}
\end{figure}
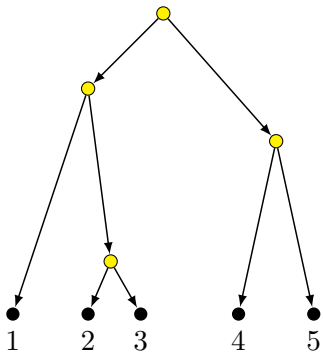

In its
simplest form each variable on the tree is binary and represents the
presence or absence of a characteristic in a large class of genetic
locations, believed to occur at random with a particular probability
determined by the vertex. The usual evolutionary hypotheses tell us that, if
the tree is valid, then collections of variables separated from each other by
any internal vertex are independent of each other given the value of that
vertex. Our problem is to select an evolutionary tree that gives the most
plausible evolutionary explanation of the data we have observed on the
extant species. For any tree there is a formal Bayesian
selection method to do this. We simply assign a prior density to each
parameter, using methods such as those described by \cite{heckerman_tutorial} respecting parameter modularity, calculate the corresponding log marginal likelihood score marginalising over the hidden variables and choose
the tree scoring the highest.

The problem occurs because it can be proved (see for example \cite{settimi2000gma}) that a tree is statistically the same collection of sample densities over observed vertices of the graph  as a simpler tree if it contains an interior vertex
with only two neighbours. A common procedure is therefore to restrict the
class of tree models to include only trees whose interior vertices have 3 or
more neighbours. However, if the ensuing inference were to depend on
the associated hypothesis that a hereditary ancestor existed \emph{only} if it
had at least two non-extinct associated species is surely not satisfactory for
two reasons. First, it seems quite conceivable that, in fact, there is only one
direct descendant still surviving from a particular species. Second, even if
we accepted the above, from an inferential point of view we should note that the property of having no degree two vertices (i.e. with two neighbours) is not closed under taking margins over a subset of the set of leaves of a tree. By this we mean that a subtree of the original tree will typically have degree two vertices. 

The fact that a marginal tree contains degree two vertices has important practical implications whenever we want to include an additional species in our data set. This happens for example in the procedure of \textit{outgroup rooting} when an outgroup is added in phylogenetic analysis in order to find the root of an unrooted phylogenetic tree as described by \cite[Section 3.1.1.2]{yang2006cme}. For a simple example imagine that the undirected tree on the left of Figure \ref{fig:outgroup} has been chosen for four extant species. To find the root of this tree an outgroup $5$ has been added. Let say  the tree on the right side of Figure \ref{fig:outgroup} has been found for the augmented data set. We have now introduced an additional hidden vertex $c$ between $3$ and $4$. In this illustrated case the marginal model over $\{1,2,3,4\}$ in the tree on the right-hand side coincides with the original model on the left. However, as we show in this paper, if we use automatic model selection methods, it is not clear that the best marginal model for $\{1,2,3,4\}$ in the augmented data set will be the original model; and this is the minimal requirement for the outgroup rooting method to work in a consistent way.

\tikzstyle{leaf}=[circle,fill=black,minimum size=5pt,inner sep=0pt] %
\tikzstyle{vertex}=[circle,fill=yellow,draw,minimum size=5pt,inner sep=0pt] 
\begin{figure}[h]
\centering
\begin{tikzpicture}
    \node[leaf] (1) at (-1,.4)  [label=above:$1$] {};
    \node[leaf] (2) at (-1,-1)  [label=below:$2$] {};
    \node[leaf] (3) at (1,1)  [label=above:$3$] {};
    \node[leaf] (4) at (2,-1)  [label=below:$4$] {};
    \node[vertex] (a) at (-.5,0) [label=above:$a$]{};
    \node[vertex] (b) at (.5,0) [label=above:$b$]{};
    \draw (a) to (1);
    \draw (a) to (2);
    \draw (a) to (b);
    \draw (3) to (b);
    \draw (4) to (b);
  \end{tikzpicture}\qquad\qquad
  \begin{tikzpicture}
    \node[leaf] (1) at (-1,.4)  [label=above:$1$] {};
    \node[leaf] (2) at (-1,-1)  [label=below:$2$] {};
    \node[leaf] (3) at (1,1)  [label=above:$3$] {};
    \node[leaf] (4) at (2,-1)  [label=below:$4$] {};
    \node[leaf] (5) at (5,.5)  [label=below:$5$] {};
    \node[vertex] (a) at (-.5,0) [label=above:$a$]{};
    \node[vertex] (b) at (.5,0) [label=above:$b$]{};
    \node[vertex] (c) at (1,-0.4) [label=above:$c$]{};
    \draw (a) to (1);
    \draw (a) to (2);
    \draw (a) to (b);
    \draw (3) to (b);
    \draw (b) to (c);
    \draw (4) to (c);
    \draw (5) to (c);
  \end{tikzpicture}
\caption{Including an outgroup may distort the analysis.}
\label{fig:outgroup}
\end{figure}
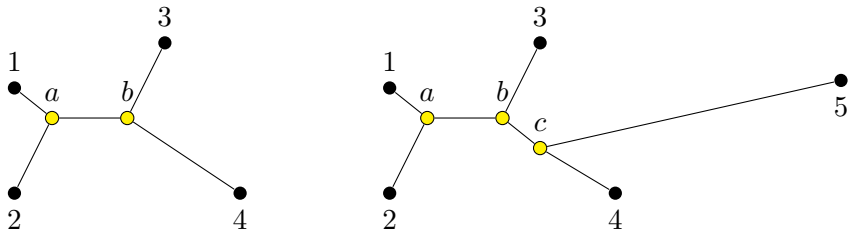

In examples like the one given by phylogenetic tree models both standard BF and also BIC model
selection methods do not in general respect the IIA property. In this paper
we examine this phenomenon in much more detail with reference to the
simplest possible manifestation of this ambiguity in Section \ref%
{sec:example}. A natural question to ask is then whether these model
selection methods are at least \emph{approximately }invariant to this choice of
embedding class (as appears to be the case  of \cite{consonni2011} when
addressing a rather different issue). Sadly the answer to this question is no!
The embedding class can have a critical impact on the model selection even
in the simplest cases. In Section \ref{sec:finite} we analyse our basic example in a full Bayes factor model selection with conjugate priors, which is a
default method in numerous \texttt{R} packages and {Hugin} software. In
Section \ref{sec:asym} we also show that the parametrization ambiguity also
applies to BIC model selection and provide an in-depth geometrical
explanation of this phenomenon. We show again that the embedding class can
have a critical impact on the model selection. We end the paper with a short
discussion of the more general implications of this phenomenon.

\section{The running example}\label{sec:example}

In our discussion we first want to make a clear distinction between different notions of statistical models used in this paper. A \textit{model} for a random variable $\mathbf{X}$ is a family of distributions of $\mathbf{X}$. A \textit{parametric model} is a parametric family $M_\psi$ of probability distributions of $\mathbf{X}$ together with the defining parametrization $\psi$, which is a map from the given finite dimensional parameter space $\Theta$ to the space of all probability distributions. We say that a model $M$ has a parametric formulation if there exists a parametrization defining this model. Let $M_{\phi}$, $M_{\psi}$ be two parametric models with parametrisations $\phi$, $\psi$ and parameter sets $\Theta$, $\Omega$ respectively. We say that $M_{\phi}$ and $M_{\psi}$ are {parametric formulations of the same model} if $\phi(\Theta)=\psi(\Omega)$ as models. Finally, a \textit{Bayesian model} is a parametric model together with an associated family of prior distributions on the parameter space $\Theta$. Hence the same model can have many parameteric formulations and each, when combined with the associated prior distribution over parameters, can lead to a different Bayesian model. 

To demonstrate the problem described in the introduction we will focus on a comparison of two simple
models for a vector of two binary random variables $X$ and $Z$ under two
different parametric formulations. The first model is the saturated model and the second
is the model of independence $X{\;\bot\!\!\!\!\!\!\bot\;} Z$. The motivation
is as follows. Suppose that we have two hypotheses: $H_{0}$ that $X$ and $Z$
are unrelated ($X{\;\bot\!\!\!\!\!\!\bot\;} Z$) and $H_{1}$ that there are
evolutionary related ($X \rightarrow Z$). What we will demonstrate is that -
with routine model settings - the second hypothesis must be distinguished
from the one where we have an evolutionary relationship of the form $%
X\rightarrow Y\rightarrow Z$, where the ancestor of $X$ and the predecessor
of $Z$ has not been observed. This is so even though both lead to exactly the same model for $(X,Z)$. The problem is that when
we refer to evolutionary relation we certainly mean both $X\rightarrow Z$, $%
X\rightarrow Y\rightarrow Z$ and even $X\rightarrow Y_{1}\rightarrow
Y_{2}\rightarrow Z$ simultaneously. The issue we have here is that therefore
from a Bayesian point of view, for model selection we need to add all those
intermediate vertices that might have occurred between the two observed
vertices. But how do we determine this number and why should it impinge on our
choice?

First, consider the ``natural'' parametric formulation where the
saturated model is parametrised by the joint distributions and the
independence model by the corresponding marginal distributions. Denote these parametric 
models by ${M}_{1}^{(0)}$ and ${M}_{0}^{(0)}$ respectively and the parameter spaces by $\Theta_{0}^{(0)}=[0,1]^{2}$ and 
$\Theta_{1}^{(0)}=\{\theta_{xz}\in \mathbb{R}^{4}:\, \sum_{i,j=0}^{1}\theta_{xz}(i,j)=1, \theta_{xz}(i,j)\geq 0\}$. Thus $M_{1}^{(0)}$ is given by $p_{xz}(i,j)=\theta_{xz}(i,j)$ and $M_{0}^{(0)}$ by $p_{xz}(i,j)=\theta_{x}(i)\theta_{z}(j)$ for $i,j=0,1$. The directed acyclic graphs (see \cite{lauritzen:96}) 
representing these models are given in Figure \ref{fig:mod1}. 
%\begin{figure}[h]
%\centering
%\nonumber\includegraphics[scale=0.7]{m1.pdf}
%\end{figure}
\tikzstyle{leaf}=[circle,fill=black,minimum size=5pt,inner sep=0pt] %
\tikzstyle{vertex}=[circle,fill=yellow,draw,minimum size=5pt,inner sep=0pt] 
\begin{figure}[h]
\centering
\begin{tikzpicture}
    \node[leaf] (1) at (0,1)  [label=above:$X$] {};
    \node[leaf] (2) at (1.5,1) [label=above:$Z$]{};
    \node[leaf] (3) at (0,0) [label=above:$X$]{};
    \node[leaf] (4) at (1.5,0) [label=above:$Z$] {};
    \node (M1) at (-2,1.1)
      {${M}_{1}^{(0)}:$};
      \node (M2) at (-2,0.1)
      {${M}_{0}^{(0)}:$};
    \draw[<-,-latex,line width=.3mm] (1) to (2);
  \end{tikzpicture}
\caption{The directed acyclic graphs of the saturated model and the independence model under the first
parametric formulation.}
\label{fig:mod1}
\end{figure}
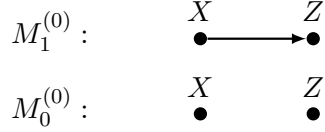

Alternatively, consider two other parametric models for $(X,Z)$: model ${M}%
^{(1)}_1$ of conditional independence $X{\;\bot\!\!\!\!\!\!\bot\;} Z|Y$, implying the saturated model on the $(X,Z)$ margin and 
represented by a graph $X\rightarrow Y \rightarrow Z$; and its submodel ${M}%
^{(1)}_0$ of marginal independence of $X$ and $Z$. Here we assume that $Y$
is binary and not observed and the model is parametrized by the marginal distribution of $X$ and conditional distributions of $Y$ given $X$ and of $Z$ given $Y$. 

More generally by $M_{1}^{(k)}$ for $k\geq 1$
denote the parametric model for $(X,Z)$ given by the graph $X\rightarrow
Y_{1}\rightarrow\cdots\rightarrow Y_{k}\rightarrow Z$, where all the $Y_{i}$
are assumed to be binary and hidden. The corresponding marginal independence model is
denoted by $M_{0}^{(k)}$ and it is a submodel of $M_{1}^{(k)}$. The parameter spaces are denoted by $\Theta_{0}^{(k)}$ and $\Theta_{1}^{(k)}$. The parameters of $M_{1}^{(k)}$ are given by the marginal distribution of $X$ and conditional probability for each arrow of the corresponding graph. There are exactly $1+2(k+1)$ free parameters denoted by $\theta_{x}(1)$, $\theta_{1|x}(1|i)$, $\theta_{2|1}(1|i)$, \ldots, $\theta_{z|k}(1|i)$ for $i=0,1$ where for example $\theta_x(1)=P(X=1)$, $\theta_{1|x}(1|0)=P(Y_1=1|X=0)$ and $\theta_{j|j-1}(1|0)=P(Y_{j}=1|Y_{j-1}=0)$. It follows that  $\Theta_{1}^{(k)}=[0,1]^{1+2(k+1)}$.

\begin{figure}[h]
\centering
\begin{tikzpicture}
    \node[leaf] (1) at (0,1)  [label=above:$X$] {};
    \node[vertex] (2) at (1.5,1) [label=above:$Y$]{};
     \node[leaf] (3) at (3,1) [label=above:$Z$]{};   
    \node[leaf] (4) at (0,0) [label=above:$X$]{};
    \node[vertex] (5) at (1.5,0) [label=above:$Y$] {};
    \node[leaf] (6) at (3,0) [label=above:$Z$] {};
    \node[leaf] (7) at (0,-1) [label=above:$X$]{};
    \node[vertex] (8) at (1.5,-1) [label=above:$Y$] {};
    \node[leaf] (9) at (3,-1) [label=above:$Z$] {};
    \node (M1) at (-2,1.1)
      {${M}_{1}^{(1)}:$};
      \node (M2) at (-2,0.1)
      {${M}_{0}^{(1)}:$};
    \draw[<-,-latex,line width=.3mm] (1) to (2);
    \draw[<-,-latex,line width=.3mm] (2) to (3);
    \draw[<-,-latex,line width=.3mm] (4) -> (5);
       \draw[<-,-latex,line width=.3mm] (8) -> (9);
  \end{tikzpicture}
\caption{The saturated model and the independence model under the second
parametric formulation with $k=1$.}
\label{fig:mod2}
\end{figure}
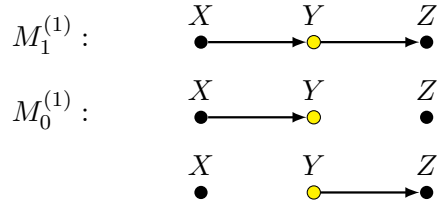

By \cite[Corollary 1]{gilula1979svd}, ${M}_1^{(k)}$ for every $k\geq 1$ is the
saturated model and hence it is equivalent to ${M}^{(0)}_1$. Since we
also have that ${M}^{(k)}_0$ is equivalent to ${M}^{(0)}_0$  then for every $k\geq 0$ we
compare the same models. Although  the models $M_{0}^{(k)}$ are the same, as parametric or Bayesian models they are very different. If $k=1$ then $M^{(1)}_{0}$ is a
union of two parametric submodels $Y\indep Z$, $X\indep Y$ of $M^{(1)}_{1}$ as depicted in Figure \ref{fig:mod2} and hence $\Theta_{0}^{(1)}$ is isomorphic to a subspace of $\Theta_{1}^{(1)}$ given as a union of two intersecting components given by equations: $\theta_{1|x}(1|0)-\theta_{1|x}(1|1)=0$ and $\theta_{z|1}(1|0)-\theta_{z|1}(1|1)=0$. The common intersection locus is the singularity of $\Theta_{0}^{(1)}$. 
 More generally, the larger is $k$ the more complicated and more \textit{singular} is the embedding of the parametric model 
$M_{0}^{(k)}$ in $M_{1}^{(k)}$. This gives the geometric intuition why the model selection for large $k$ may differ from $k=0$.

A statistical reason why IIA may fail follows from the discussion in the introduction. Thus, for fixed $k\geq 0$ consider the class $\mathbb{M}_{k}$ of Bayesian models $M_{1}^{(k)}$
and $M_{0}^{(k)}$ together with the associated families of all prior
distributions over the corresponding parameter spaces. First note that $\mathbb{M}_{0}$ can be naturally embedded into $\mathbb{M}_{1}$ because $M_{0}^{(0)}=M_{0}^{(1)}$ and $M_{1}^{(0)}=M_{1}^{(1)}$ as models and the parameter space $\Theta_{1}^{(0)}$ can be associated with a subspace of $\Theta_{1}^{(1)}$ for which $X$ and $Y$ are functionally related (e.g. $\theta_{y|x}(1|1)=\theta_{y|x}(0|0)=1$). Hence every prior on $\Theta_{1}^{(0)}$ can be treated as a prior concentrated on this subspace. There is no statistical way of distinguishing between $M_{1}^{(0)}$ and its copy embedded in $M_{1}^{(1)}$.  More generally 
$\mathbb{M}_{k-1}\subset \mathbb{M}_{k}$ for all $k\geq 1$. This follows from the fact that every prior distribution for parameters of the Bayesian models in $\mathbb{M}_{k-1}$ is a degenerate prior distribution of $\mathbb{M}_{k}$. Hence, at least in principle, the statistical analysis should not dependent on a particular embedding. 

The problem is that the
containment $\mathbb{M}_{k-1}\subset \mathbb{M}_{k}$ breaks down whenever we assume some sort of regularity of the
prior distributions ruling out some possible Bayesian models. In this case the analysis may highly depend on $k$. This is particularly evident in the case of the BIC criterion which is derived under the assumption that prior distributions are diffuse (bounded and bounded away from zero) and hence then cannot be degenerate. The problem with diffuse priors and the BIC criterion has been reported in other contexts. In particular it has been shown that BIC tends to support simpler models which follows from the Jeffrey-Lindley's paradox (see for example \cite{denison2002bayesian}). Recently it has been shown by  \cite{johnson2010nonlocal} and   \cite{consonni2011}, that using local priors causes in particular a very slow convergence to the true smaller model. In Section \ref{sec:asymptotic} we show that this problem may be particularly important when the choice of the parametric formulation is not clear. 

\section{Finite sample conjugate selection}\label{sec:finite}

Perhaps the most common way to set up the prior densities across a large
class of graphical models is to ensure each decomposable model within the class has 
consistent priors and the same strength in the following sense. Let $\pi=({
\pi}_{\mathbf{x}})$ denote the vector of prior probabilities that a unit from the model
population takes level $\mathbf{x}\in \mathcal{X}$ where $\mathcal{X}$ is
the sample space of the problem. This joint prior distribution induces marginal and conditional distributions over subvectors of $\mathbf{x}$.

\begin{itemize}
\item Define $\boldsymbol{\beta}_{\mathbf{x}}=\beta {\pi }_{\mathbf{x}}$ for $\mathbf{x}\in \mathcal{X}$, where the
positive real scalar $\beta$, called the \textit{effective sample size}, reflects the number of observations the modeler
believes her prior beliefs are worth. 

\item Differentiate each model in the class with the
appropriate hyper-Dirichlet prior (see \cite%
{dawidlauritzen}) faithful to its particular sets of
conditional independence assumptions defined by its graph. The hyper-Dirichlet priors form the conjugate class for decomposable graphical models. They are widely used in numerous 
\texttt{R} packages and in Hugin (see \cite%
{springerlink:10.1007/978-3-540-45062-7_49}).

\item Ensure the relevant prior  expected clique cell probabilities and
effective sample size parameters of these different product Dirichlets are
consistent across the different models in the class. This is obtained  by choosing the
associated Dirichlet parameters $\boldsymbol{\beta }_{i}$ of the different (marginal or conditional) components $C_i$ of
BN model with prior probability $\pi_i$ so that $\boldsymbol{\beta}_i=\beta\pi_i$. In this way these Dirichlet distributions over various components are consistent with those of a Dirichlet over a saturated model with parameters  $\boldsymbol{\beta }_{\mathbf{x}}$ for  $\mathbf{x}
\in \mathcal{X}$. This in particular assures that the parameter modularity holds.
\end{itemize}

In what follows we therefore faithfully follow this procedure applying it to
the different embeddings below, in addition ensuring that different
embedding match in an analogous way. In this section we show that the selection procedure between these Bayesian models will not satisfy the IIA. 

First we compare the models for $k=0,1$. Let $\pi_{xyz}$ be the joint prior distribution on $(X,Y,Z)$. By $\pi_{xz}$, $\pi_x$, $\pi_z$, $\pi_{y|x}$ and $\pi_{z|y}$ we denote the corresponding prior marginal and conditional probabilities. Let first $k=0$. For ${M}^{(0)}_1$ we set the prior distribution to be the Dirichlet distribution, $\theta_{xz}\sim\mathrm{Dir}(
\boldsymbol{\beta}_{xz})$, where $\boldsymbol{\beta}_{xz}=\beta \pi_{xz}$ and $\beta>0$. For ${M}^{(0)}_0$ we have $\theta_{x}\sim \mathrm{Beta}(\boldsymbol{\beta}_{x})$ and 
$\theta_{z}\sim \mathrm{Beta}(\boldsymbol{\beta}_{z})$, where $\boldsymbol{\beta}_x=\beta\pi_x$ and $\boldsymbol{\beta}_z=\beta\pi_z$. Let now $k= 1$. The standard conjugate analysis for $M_1^{(1)}$ requires setting hyper-Dirichlet priors for the joint distribution of $(X,Y,Z)$. We set $\theta_{x}\sim \mathrm{Beta}(\boldsymbol{\beta}_{x})$, $
\theta_{y|x}(1|i)\sim \mathrm{Beta}(\boldsymbol{\beta}_{y|x}(i))$  and $
\theta_{z|y}(\cdot|i)\sim \mathrm{Beta}(\boldsymbol{\beta}_{z|y}(i))$, where $\boldsymbol{\beta}_{y|x}(i)=\beta\pi_{y|x}(\cdot|i)$ and $\boldsymbol{\beta}_{z|y}(i)=\beta\pi_{z|y}(\cdot|i)$ for $i=0,1$. If we
assume that all five random variables $\theta_x$, $\theta_{y|x}(\cdot|0)$, $\theta_{y|x}(\cdot|1)$, $\theta_{z|y}(\cdot|0)$ and $\theta_{z|y}(\cdot|1)$ are independent, then the variable $p_{xyz}$, where $p_{xyz}(i,j,k)=\theta_{x}(i)\theta_{y|x}(j|i)\theta_{z|y}(k|j) $, has a hyper-Dirichlet
distribution. This induces a distribution  of $p_{xz}$ by
\begin{equation}  \label{eq:pik}
p_{xz}(i,k)=p_{xyz}(i,0,k)+p_{xyz}(i,1,k)=\theta_{x}(i)\sum_{j=0}^{1} \theta_{y|x}(j|i)\theta_{z|y}(k|j)
\end{equation}
which is \textit{not} in general Dirichlet as for $M_1^{(0)}$ and so gives a different value of the marginal likelihood in this case. 
   
Since the induced distribution of $p_x$ and $p_z$ is the same as for $k=0$, the difference in the analyses performed for the Bayes factor follows from the difference in the marginal likelihoods of the
saturated models. For $M_0^{(0)}$ and $M_1^{(0)}$ we can easily obtain
formulae for the posterior distribution and for the marginal likelihood
functions of the sample counts $\mathbf{u}=[u_{ij}]$. So the Bayes factor
can be calculated directly. For any table $\alpha=[\alpha_{i}]$ define the Beta function $B(\alpha)=\frac{\prod_{i}
\Gamma(\alpha_{i})}{\Gamma(\sum_{i}\alpha_{i})}$ where $\Gamma(\cdot)$ is the Gamma function. The marginal likelihood for ${M}_{1}^{(0)}$ is: {\ 
\begin{eqnarray}  \label{eq:ML11}
L_1^{(0)}\quad=\quad\frac{B(\boldsymbol{\beta}_{xz}+\mathbf{u})}{B(\boldsymbol{\beta}_{xz})},
\end{eqnarray}
} The marginal likelihood for ${M}%
_{0}^{(0)}$ is: {\ 
\begin{eqnarray}  \label{eq:ML10}
L_0^{(0)}\quad=\quad\frac{B(\boldsymbol{\beta}_{x}+\mathbf{u}_{x})}{B(\boldsymbol{\beta}_{x})}\frac{B(\boldsymbol{\beta}_{z}+\mathbf{u}_{z})}{B(\boldsymbol{\beta}_{z})},
\end{eqnarray}
}
where  $\mathbf{u}_{x}$ and $\mathbf{u}_{z}$ denote the marginal counts for $X$ and $Z$ respectively.

To obtain the marginal likelihood $L_1^{(1)}$ of $M_1^{(1)}$ the prior density for $p_{xz}$ can be written explicitly
using (\ref{eq:pik}). Exact computations are technically inelegant but some recent developments of \cite{lin2009marginal} make it possible to compute the marginal integrals exactly at least for simple mixture models. We note that
simple Monte Carlo approximations of the integrals below give very good
results as well.

As an example consider the following table of counts: 
\begin{equation*}
\mathbf{u}=\left[%
\begin{array}{cc}
13 & 9 \\ 
4 & 18%
\end{array}
\right].
\end{equation*}
The exact Fisher's test in this case gives the value of the odds ratio $6.2$ and the corresponding $p$-value is $0.012$. Hence the data strongly support the alternative hypothesis. The Bayes factor for the first parametric formulation, denoted by ${\rm BF}_{0}$, is easily computed by dividing $L_1^{(0)}$ in  (\ref%
{eq:ML11}) by $L_0^{(0)}$ in (\ref{eq:ML10}). For $k=1$ we calculated scores using  \cite{lin2009marginal} confirming these against good approximate results provided by simple Monte Carlo simulation. Initially assume that $\beta=4$ and $
\pi_{xyz}(i,j,k)=1/8$ for all $i,j,k=0,1$. This gives $
p_{xz}$ a Dirichlet ${\rm Dir}(1,1,1,1)$ (and hence uniform) distribution if $k=0$ (but not if $k=1$) and: 
\begin{equation}\label{eq:BFunif}
\mathrm{BF}_{0}=11.47,\quad \mathrm{BF}_{1}=6.27.
\end{equation}
So in the second parametric formulation, when $k=1$, the Bayes factor slightly underestimates the evidence for
the saturated model. The reason here is that the induced prior distribution
on $p_{y}$, where $p_{y}(j)=\sum_{i,k}p_{xyz}(i,j,k)$, is not uniform and equals $\mathrm{Beta}(2,2)$. This is an important point because it shows that the uniform prior $\pi$ may lead to highly informative scenarios, which may then affect our analysis. 

\begin{figure}[htp!]
\begin{center}
\includegraphics[scale=0.6]{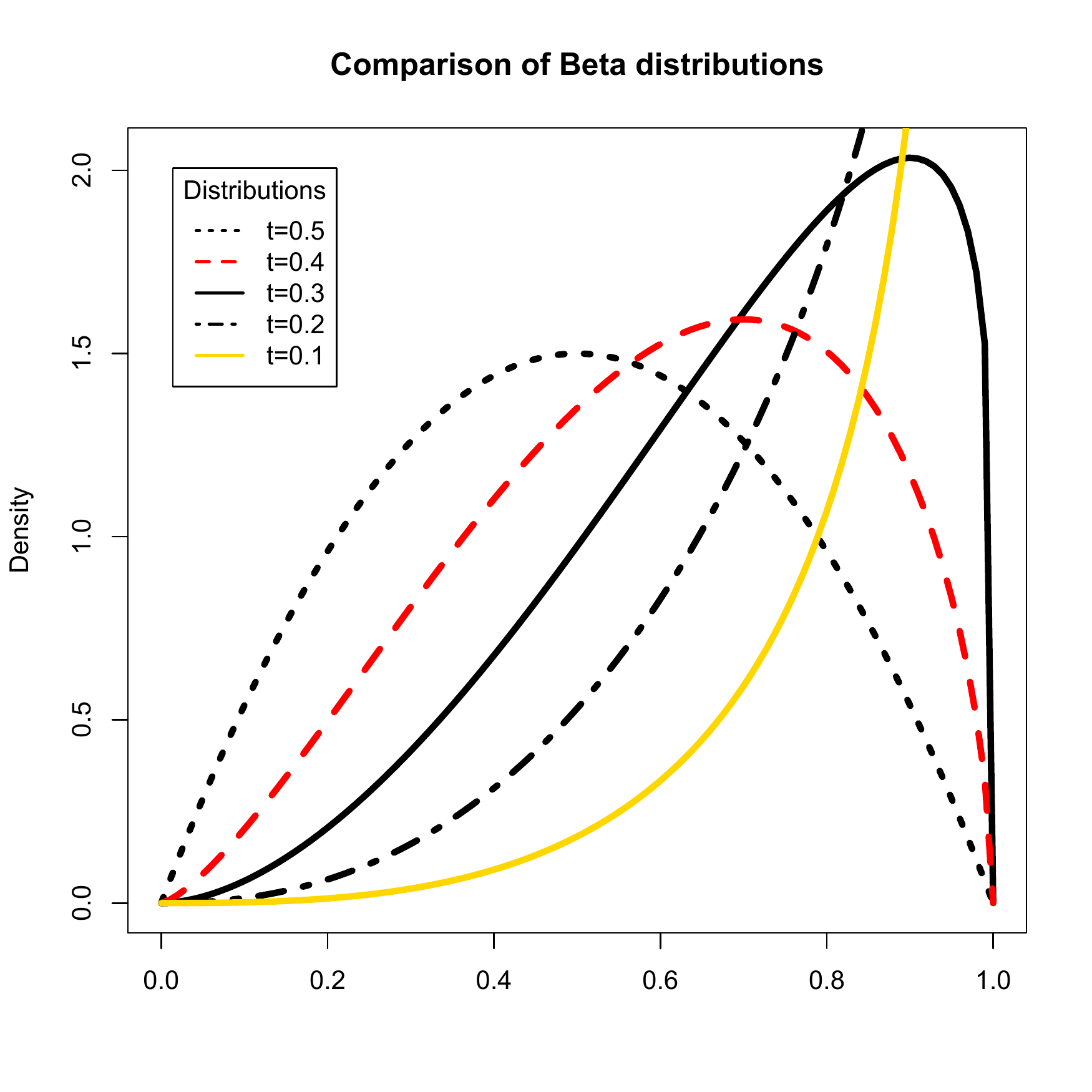}
\end{center}
\caption{The densities of Beta distributions with parameters $(4(1-t),4t)$, where $t=0.5,0.4,0.3,0.2,0.1$. }\label{fig:betas}
\end{figure}

The result of
the analysis changes if we set $\beta=4$ and $\pi_{xyz}(i,0,k)=t/4$ and $%
\pi_{xyz}({i,1,k})=(1-t)/4$ for $t<0.5$. Note that the induced distribution on $p_{xz}$ in the first
parametric formulation is still uniform because $\pi_{xz}({i,k})=1/4$ for all $i,k=0,1$ but the prior information on the distribution of
the hidden variable is much stronger. This affects the distribution of $%
p_{xz}$ under the second parametric formulation. The induced distribution of $p_{y}$ is Beta with parameters $(4(1-t),4t)$. In particular, for $t\leq 0.25$ the corresponding density is not bounded giving increasingly more probability to the event $p_{y}(1)>1-\epsilon$ as $t\rightarrow 0$. Some Beta density functions for different values of $t$ are given in Figure \ref{fig:betas}.

If $t$ is small then there is a strong a priori
information that the inner vertex is close to being degenerate. Therefore, the Bayesian 
model $M_1^{(1)}$ is close to the model of independence. This should cause
overestimation of the independence model. Indeed, if $t=0.2$ then  for $\mathbf{u%
}$ given above we have: 
\begin{equation*}
\mathrm{BF}_{0}=11.47,\quad \mathrm{BF}_{1}=2.68.
\end{equation*}

This example illustrates a more serious problem. For each $t$, as above, we compute $\mathrm{BF}%
_{0}(t)$ and $\mathrm{BF}_{1}(t)$ obtaining the following result. 
\begin{proposition}\label{prop:1}
For every table $\mathbf{u}$ the Bayes Factor ${\rm BF}_{0}(t)$ is constant and does not depend on $t$. Moreover, for every $\mathbf{u}$, ${\rm BF}_{1}(t)\rightarrow 0$ almost surely as $t\rightarrow 0$.
\end{proposition}

This result may seem obvious. However, what it really shows is that, even though the induced priors on the joint distribution of $%
(X,Z)$ follow from the same joint prior distribution $\pi$ on $(X,Y,Z)$, in the first case the Bayes factor may give evidence in
favour of the saturated model and the second in favour of the model of
independence. 

The situation becomes even more dramatic if $k$ increases. Thus if $k=2$ we set $\beta=4$ and $\pi_{x12z}(i,j,k,l)=1/16$ for all $i,j,k,l=0,1$ and again the induced prior on $p_{xz}$ for the case $k=0$ is uniform. Under this setting we find that ${\rm BF}_{2}=2.13$ which is to be compared with (\ref{eq:BFunif}). This uniform case is easy to generalize and illustrates another serious issue related with the choice of the parametric formulation for the model under consideration. 

For general $k\geq 1$, if  $\beta=4$ and $\pi_{x 1\cdots k z}(\mathbf{i})=2^{-(k+2)}$ for every  $\mathbf{i}=(i_0,\ldots,i_{k+1})\in \{0,1\}^{k+2}$, then the induced prior distributions are $\theta_x\sim {\rm Beta}(2,2)$, $\theta_{1|x}(\cdot|i),\ldots, \theta_{z|k}(\cdot|i)\sim {\rm Beta}(1,1)$ for $i=0,1$. 
\begin{proposition}\label{prop:2}
Let $\beta=4$ and  $\pi_{x 1\cdots k z}(\mathbf{i})=2^{-(k+2)}$ for every $\mathbf{i}\in \{0,1\}^{k+2}$. Then ${\rm BF}_{k}\rightarrow 0$ as $k\rightarrow \infty$. 
\end{proposition}
The proof is given in Section \ref{app:prop2}.

In the case of Proposition \ref{prop:1} we analysed the discrepancy of the model selection if the prior of the hidden variable becomes degenerate without affecting the prior over the observed variables. In Proposition \ref{prop:2} however the prior distribution of every hidden variable is symmetric and hence it shows a different aspect of the discussed problem.  

\section{Asymptotic analysis}\label{sec:asymptotic}\label{sec:asym}

Perhaps not surprisingly a similar ambiguity also arises if we use the popular BIC rather than Bayes Factor model selection method. Let $q$ denote the true density of $(X,Z)$, assumed strictly positive everywhere,  and let 
$(X_{1},Z_{1}),\ldots,(X_{n},Z_{n})$ be a random sample from this
distribution. In this section we perform an asymptotic analysis. Let $\mathcal{Z}_{n}$ denote the marginal likelihood
function and $S_{q}$ the entropy function of $q$. We define $F_{n}=-\log \mathcal{Z}_{n}$.

In the natural parametrisation the asymptotic likelihood as 
$n\rightarrow \infty$ is always maximised over the unique point $q$. By the
result of \cite{schwarz1978edm} the asymptotic formula for the marginal
likelihood can be obtained using the Laplace approximation. Whenever the prior distributions are bounded and bounded away from zero then, as $n\rightarrow
\infty$, 
\begin{equation}  \label{eq:bic}
\mathbb{E}F_{n}=n S_{q} - \frac{d}{2}\log n + O(1),
\end{equation}
where $d=2$ for the marginal likelihood given the model ${M}_{0}^{(0)}$ and $d=3$ for model ${M}_{1}^{(0)}$. This asymptotic approximation of the marginal likelihood justifies the use of the well known BIC penalty used in routine model selection. If the true distribution lies in the null model then the entropies  $S_{q}$ for both models will be asymptotically equal and the difference in scores ${\rm BIC}_{0}-{\rm BIC}_{1}=\frac{1}{2}\log n$ will be always positive, which gives a positive evidence in favour of the null model. 

The interpretation of BIC in the presence of hidden variables is more subtle. Under the second proposed parametrisation $k=1$ both models are
unidentifiable hence the asymptotic analysis is much harder. In particular the Laplace approximation is no longer formally valid and the appropriate asymptotic analysis must use the results of singular learning theory developed by \cite{watanabe_book}. 

We now compute the marginal likelihood of the data under the parametric model $M_{1}^{(k)}$ for $k\geq 1$. The asymptotic formula depends on the true distribution $q$ generating the data. If $q\in M_{1}^{(k)}\setminus M_{0}^{(k)}$ for $k\geq 1$ then, despite the identifiability issue, the correct asymptotic approximations can be shown to be equal to the classical BIC formula in the case when $k=0$. The problem occurs when $q\in M_{0}^{(k)}$. In this case the set of parameters mapping to  $q$  is highly singular. Whenever the prior distribution is
bounded and bounded away from zero on the whole parameter space, the asymptotic
approximation of $\mathbb{E}F_{n}$ for the model $M_{1}^{(k)}$, as $n\rightarrow \infty$, is
\begin{equation}  \label{eq:bic_sing}
\mathbb{E}F_{n}=\left\{\begin{array}{ll}
n S_{q} - \frac{3}{2}\log n + O(1) & \mbox{if }q\in M_{1}^{(k)}\setminus M_{0}^{(k)},\\
n S_{q} - \frac{3}{2}\log n + k\log \log n + O(1) & \mbox{if }q\in M_{0}^{(k)}.
\end{array}\right.
\end{equation}
The proof of (\ref{eq:bic_sing}) is given in Section \ref{app:proofeq}. 

\begin{figure}[htp!]
\begin{center}
\includegraphics[scale=0.6]{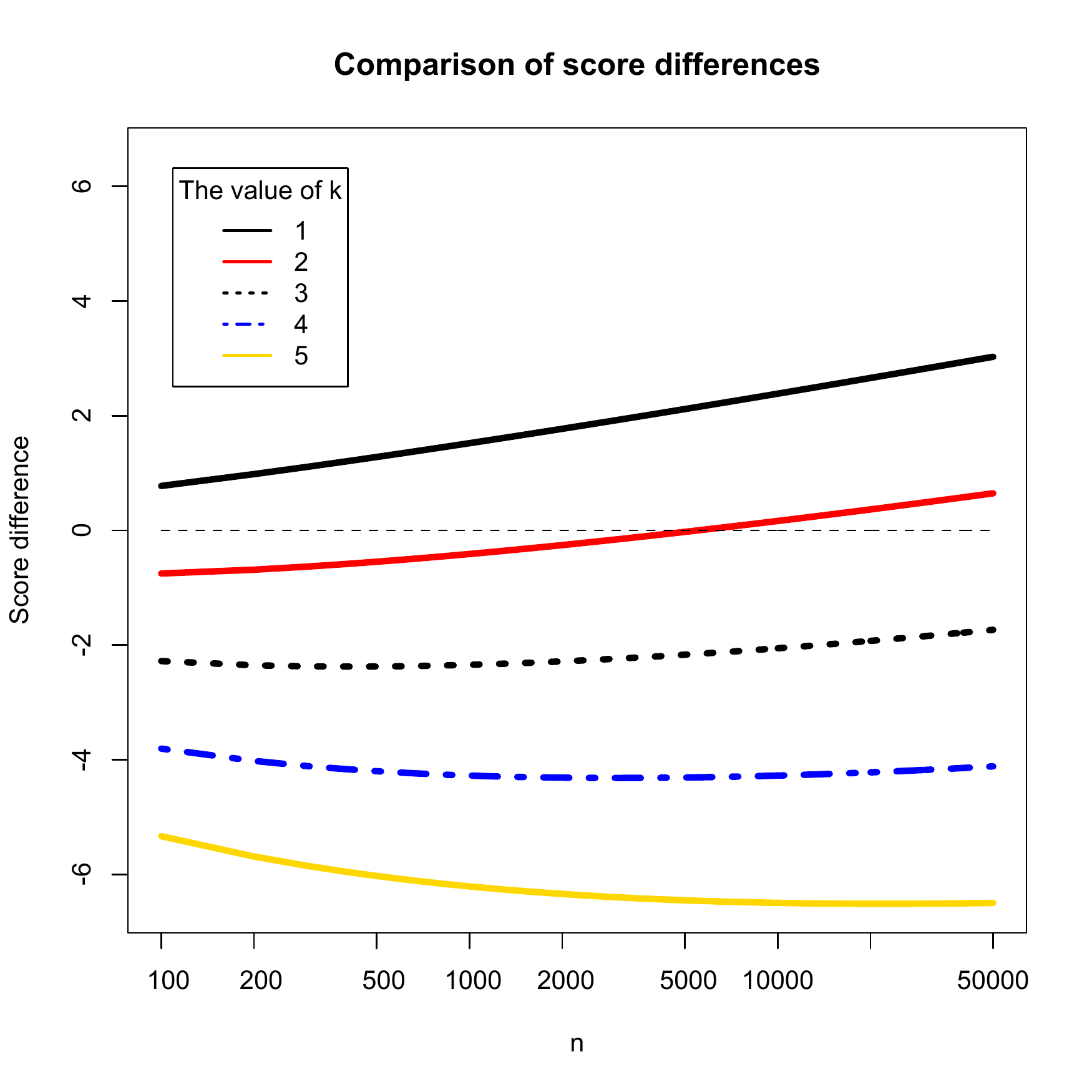}
\end{center}
\caption{The difference in scores $\frac{1}{2}\log n-k\log\log n$ depicted as a function of $n$ for different values of $k=1,2,3,4,5$. The $n$ axis is in the log scale.}\label{fig:scores}
\end{figure}

The marginal likelihood of $M_{0}^{(k)}$ is equal to the marginal likelihood of $M_0^{(0)}$ and thus in this case
$$
\mathbb{E}F_{n}=n S_{q} - \log n + O(1).$$ 
Now we see that a problem might occur if the true data generating distribution lies in the independence model. The bigger $k$, the harder it gets to distinguish $M_{0}^{(k)}$ from $M_{1}^{(k)}$. Since the entropy value will be asymptotically the same in both cases, the difference in scores between those two models is 
$$
(-\log n)-(-\frac{3}{2}\log n+k \log\log n)=\frac{1}{2}\log n-k\log\log n.
$$
Since the true model is the model of independence we expect this difference to be highly positive like in the case when $k=0$. However, the component of the penalty $k \log\log n$ distorts this as is shown in Figure \ref{fig:scores}. The score difference remains negative whenever $k\geq 2$ even for very large $n$. Hence, for all usual values of the sample size $n$ the use of the standard BIC criterion  underestimates the evidence of the null model whenever $k\geq 1$. 

\section{Discussion and conclusions}

In this paper we have shown that we need a foundation for justifying a particular embedding before BF or BIC model selection is unambiguous for binary tree models. This fact can be shown also to apply to all model selection over discrete graphical
models, Gaussian graphical models with potential hidden variables and, in particular, Bayesian hierarchical models, where systematically hidden variables are routinely added to the system to articulate various types of dependence structures. Problems are particularly acute when the dimension of
the embedding class is itself contextualy ambiguous as in our running
example. For all these models whenever the appropriate embedding is not
transparent, we should be aware this choice could be critical to the result
of our selection. In particular the current praxis of paying little
attention to the different inferential implications of a chosen embedding,
focusing instead on the numerical efficacy of a particular representation is
one that should be of great concern to Bayesian modelers. We note that the
problems we identify here do not concern just BF or BIC: other Bayesian
model selection methods also suffer the same difficulty.

What can we do to address this issue? First, if there exists an associated
meaning to a given embedding then we could elicit a prior distribution for
each model and then average over this. In our example we could for example
elicit the number of potential differently evolved ancestors for which only
one direct descendant survived. However, as in our example, it may well be
difficult to unambiguously make this association and is certainly not in the
spirit of current model selection, which tries fot the sake of "objectivity"
to avoid the incorporation of as little domain knowledge as possible:
including much more direct domain knowledge than this. A second possible
direction is to systematically check the plausibility of a particular
embedding on the associated marginal likelihood of different models to check
how the system will learn in various contingencies and calibrate to that.

Finally we could question, as some others do e.g. Dennis Lindley, \cite{draper1999bayesian} whether model selection is actually compatible with Bayesian
methodology at all. It would be sad however to discard these techniques
which have undoubtedly provided such useful output to scientists
endeavouring to understand the processes underpinning their observations.
But at least when using Bayesian model selection techniques in conjunction
with apparently innocuous homeneiety assumptions like, in the case of BN
model selection, parameter modularity we should be aware that the associated
inferences could seriously mislead the investigator. 

\appendix

\section{Appendix}

\subsection{Proof of Proposition \ref{prop:2}}\label{app:prop2}

Since the marginal likelihood $L^{(k)}_0$ for the null model does not change with $k$, it suffices to show that the marginal likelihood $L_{1}^{(k)}$ of the model $M_{1}^{(k)}$ converges to zero as $k\rightarrow \infty$. Let $\boldsymbol{\theta}=(\theta_{x},\theta_{1|x},\theta_{2|1},\cdots,\theta_{z|k})\in [0,1]^{1+2(k+1)}$ and  
\begin{equation*}
f_{k}(\boldsymbol{\theta})\quad=\quad\prod_{i,j=0}^{1}\left(\theta_{x}(i)\sum_{i_{1},\ldots,i_{k}}\theta_{1|x}(i_{1}|i)\theta_{2|1}(i_{2}|i_{1})\cdots \theta_{k|k-1}(i_{k}|i_{k-1})\theta_{z|k}(j|i_{k})\right)^{u_{ij}}.
\end{equation*}
Since $\theta_x\sim {\rm Beta}(2,2)$, $\theta_{1|x}(\cdot|i),\ldots, \theta_{z|k}(\cdot|i)\sim {\rm Beta}(1,1)$ for $i=0,1$, we have
\begin{equation}\label{eq:Lk}
L_{1}^{(k)}\quad=\quad \frac{\Gamma(4)}{\Gamma(2)^2}\int_{[0,1]^{1+2(k+1)}} f_{k}(\boldsymbol{\theta})\prod_{i=0}^{1}\theta_{x}(i) {\rm d}\boldsymbol{\theta}.
\end{equation}

Let $t^i_{i_1\cdots i_k}= \theta_{1|x}(i_{1}|i)\theta_{2|1}(i_{2}|i_{1})\cdots \theta_{k|k-1}(i_{k}|i_{k-1})$ for $i,i_1,\ldots,i_k=0,1$. Note that $\varphi(x)=x^a$ is a convex function on $[0,\infty)$ whenever $a\geq 1$ or $a=0$. Since $\sum_{i_{1},\ldots,i_{k}} t^i_{i_1\cdots i_k}=1$ for $i=0,1$ and either $u_{ij}\geq 1$ or $u_{ij}=0$, by Jensen's inequality
$$
\left(\theta_{x}(i)\sum_{i_{1},\ldots,i_{k}}t^i_{i_1\cdots i_k}
\theta_{z|k}(j|i_{k})\right)^{u_{ij}}\leq \sum_{i_{1},\ldots,i_{k}} t^i_{i_1\cdots i_k}(\theta_x(i)\theta_{z|k}(j|i_k))^{u_{ij}}
$$
and hence
\begin{equation}\label{eq:jensen}
f_k(\boldsymbol{\theta})\quad\leq\quad \sum_{i_1,\ldots,i_k}\prod_{i,j=0}^1\theta_{1|x}(i_{1}|i)\theta_{2|1}(i_{2}|i_{1})\cdots \theta_{k|k-1}(i_{k}|i_{k-1})\left(\theta_x(i)\theta_{z|k}(j|i_{k})\right)^{u_{ij}}.
\end{equation}
We have
$$
\int_{[0,1]^2}\prod_{i,j=0}^1\theta_{1|x}(i_1|i){\rm d}\theta_{1|x}=\int_{[0,1]^2}\theta_{1|x}(i_1|0)^2\theta_{1|x}(i_1|1)^2{\rm d}\theta_{1|x}(1|1){\rm d}\theta_{1|x}(1|0)=\left(\frac{\Gamma(3)}{\Gamma(4)}\right)^2
$$
and for $l=2,\ldots,k$ 
$$
\int_{[0,1]}\prod_{i,j=0}^1\theta_{l|l-1}(i_l|i_{l-1}){\rm d}\theta_{l|l-1}(1|i_{l-1})=\int_{[0,1]}\theta_{l|l-1}(i_l|i_{l-1})^4{\rm d}\theta_{l|l-1}(1|i_{l-1})=\left(\frac{\Gamma(5)}{\Gamma(6)}\right)^2.
$$
This together with (\ref{eq:jensen}) gives
$$
L_1^{(k)}\leq C(\mathbf{u})\sum_{i_1,\ldots,i_k} \left(\frac{\Gamma(5)}{\Gamma(6)}\right)^{k-1}\left(\frac{\Gamma(3)}{\Gamma(4)}\right)^2=C(\mathbf{u})\,\frac{5}{9}\left(\frac{2}{5}\right)^k,
$$
where $C(\mathbf{u})$ is a constant which depends only on the table of counts $\mathbf{u}$. This in particular implies that $L_1^{(k)}\rightarrow 0$ as $k\rightarrow \infty$.

\subsection{Proof of Equation (\protect\ref{eq:bic_sing})}\label{app:proofeq}

The set of parameters of the model is the marginal distribution $p_{x}$ of $X$, the conditional distribution of $Y_{1}$ given $X=i$ denoted by $\theta_{1}(\cdot|i)$, the conditional distribution of $Z$ given $Y_{k}=i$, denoted by $\theta_{z}(\cdot|i)$ and conditional distributions of $Y_{j}$ given $Y_{j-1}=i$ for every $j=2,\ldots,k$ denoted by $\theta_{j}(\cdot|i)$. Hence the parameter vector $\theta$ lies in $[0,1]^{1+2(k+1)}$. We have
$$
p_{xz}(i,k)=p_{x}(i)\sum_{j_{1},\ldots,j_{k}}\theta_{1}(j_{1}|i)\theta_{2}(j_{2}|j_{1})\cdots \theta_{k}(j_{k}|j_{k-1})\theta_{z}(k|j_{k}).
$$

Since the true distribution is assumed to lie in the independence model we
cannot use the standard Laplace approximation for the marginal likelihood
because the asymptotic likelihood is maximised over a singular subset of the parameter space. Assume that the prior distribution on $\Theta$ is bounded and bounded away from zero. Then by
The Corollary 6.1 of \cite{watanabe_book} the marginal likelihood is asymptotically, as $n\rightarrow \infty$,
approximated by 
\begin{equation*}
nS-\lambda\log n+(m-1)\log\log n +O(1), 
\end{equation*}
where $\lambda$ and $m$ are the smallest pole and its multiplicity of an
analytic function given by 
\begin{equation}\label{eq:zeta}
\xi(w)=\int_{\Theta}(f(\theta))^{-w}\mathrm{d}%
\theta,
\end{equation}
where $f(\theta)$ is the Kullback-Leibler divergence from the true model $q$. By Theorem 1.2 of \cite{shaowei_rlct} we can also replace $f(\theta)$ with the sum of squares $\sum_{i,j=0}^{1}(p_{ij}(\theta)-q_{ij})^{2}$.  

Let $\mu_{x}(\theta)=p_{10}(\theta)+p_{11}(\theta)$, $\mu_{z}(\theta)=p_{01}(\theta)+p_{11}(\theta)$ and
$$
\mu_{xz}(\theta)=p_{11}(\theta)-(p_{10}(\theta)+p_{11}(\theta))(p_{01}(\theta)+p_{11}(\theta)).
$$
It is immediate to see that $\mu_{x},\mu_{z},\mu_{xz}$ are in one-to-one correspondence with $[p_{ij}]$. Moreover, $\mu_{xz}$ is just the covariance between $X$ and $Z$ and hence it is zero if and only if $X\indep Z$. Since the pole of (\ref{eq:zeta}) and its multiplicity do not change under isomorphisms, we can further
replace the function $\sum_{i,j}((p_{ij}(\theta)-q_{ij})^{2}$ with 
\begin{equation}  \label{eq:auxf}
(\mu_{x}(\theta)-\mu^{*}_{x})^{2}+(\mu_{z}(\theta)-\mu^{*}_{z})^{2}+(%
\mu_{xz}(\theta)-\mu^{*}_{xz})^{2}.
\end{equation}
The asterisks refer to the moments of the true distribution $%
q$. Note that since $q$ lies in the independence model then $\mu_{xz}^{*}=0$.

Now make an isomorphic change of coordinates of $\theta$ to parameters $$\omega=(\mu_{x}, \mu_{y_{1}},\ldots, \mu_{y_{k}}, \mu_{z}; \eta_{xy_{1}},%
\eta_{y_{1}y_{2}},\ldots, \eta_{y_{k}z})$$ where $\mu_{y_{i}}$ is the mean of $Y_{i}$ and $\eta_{xy}=\mathbb{P}(Y=1|X=1)-\mathbb{P}(Y=1|X=0)$ is the linear
regression coefficient of $Y$ with respect to $X$. Since 
\begin{equation*}
\mu_{xz}(\omega)=\frac{1}{4}(1-\mu_{x}^{2})\eta_{xy_{1}}\cdots\eta_{y_{k}z} 
\end{equation*}
then the function in (\ref{eq:auxf}) expressed in new parameters becomes 
\begin{equation*}
(\mu_{x}-\mu^{*}_{x})^{2}+(\mu_{z}-\mu^{*}_{z})^{2}+(\frac{1}{4}%
(1-\mu_{x}^{2})\eta_{xy_{1}}\cdots\eta_{y_{k}z})^{2}. 
\end{equation*}
By Remark 7.2 of \cite{watanabe_book}, if $\omega^{*}$ is an interior point of the parameter space, $\lambda=1+\lambda'$ and $m=m'$ where $%
(\lambda',m')$ are the smallest pole and its multiplicity
of the analytic function of $w$ given by 
\begin{equation*}
\int_{[-\epsilon,\epsilon]^{k+1}}(\eta_{xy_{1}}\cdots\eta_{y_{k}z})^{-2w}%
\mathrm{d}\eta_{xy_{1}}\cdots\mathrm{d}\eta_{y_{k}z}, 
\end{equation*}
for a sufficiently small $\epsilon>0$. Finally, this integral is equal to $%
C(\epsilon)(\frac{1}{1-2w})^{k+1}$, where $C(\epsilon)$ is a constant which
depends only on $\epsilon$. It follows that the pole of this function is $\lambda'=1/2$ and the multiplicity of this pole is $%
m'=k+1$ and hence $\lambda=3/2$ and $m=k+1$.

Note that to use Remark 7.2 of \cite{watanabe_book} we assume that the parameter space has locally a product structure where $\mu_{x}$ and $\mu_{z}$ are independent of other parameters. This holds only in the interior of the parameter space so the boundary points $\omega^{*}$ need to be checked separately. We omit the details.

\section*{Acknowledgements}

We would like to thank Shaowei Lin for help with setting up the exact computations in Section \ref{sec:finite}.

%\bibliographystyle{siam}
%\bibliography{../!bibliografie/algebraic_statistics}
\end{document}